

\magnification = 1200
\overfullrule=0pt

\font\titlerm = cmr10 scaled\magstep 4
\font\titlerms = cmr7 scaled\magstep 4
\font\titlermss = cmr5 scaled\magstep 4
\font\titlei = cmmi10 scaled\magstep 4
\font\titleis = cmmi7 scaled\magstep 4
\font\titleiss = cmmi5 scaled\magstep 4
\font\titlesy = cmsy10 scaled\magstep 4
\font\titlesys = cmsy7 scaled\magstep 4
\font\titlesyss = cmsy5 scaled\magstep 4
\font\titleit = cmti10 scaled\magstep 4

\def\titlefont{\def\rm{\fam0\titlerm}
\def\it{\fam\itfam\titleit}
\textfont0 = \titlerm
\scriptfont0 = \titlerms
\scriptscriptfont0 = \titlermss
\textfont1 = \titlei
\scriptfont1 = \titleis
\scriptscriptfont1 = \titleiss
\textfont2 = \titlesy
\scriptfont2 = \titlesys
\scriptscriptfont2 = \titlesyss
\textfont\itfam = \titleit
\rm}

\def\sectionfont{\def\rm{\fam0\tenrm}
\def\it{\fam\itfam\tenit}
\def\bf{\fam\bffam\tenbf}
\textfont0 = \tenrm
\scriptfont0 = \sevenrm
\scriptscriptfont0 = \fiverm
\textfont1 = \teni
\scriptfont1 = \seveni  \scriptscriptfont1=\fivei
\textfont2 = \tensy
\scriptfont2 = \sevensy
\scriptscriptfont2 = \fivesy
\textfont\itfam = \tenit
\textfont\bffam = \tenbf
\rm}

\font\teenyfont = cmr5

\global\baselineskip = 1.2\baselineskip
\global\parskip = 4pt plus 0.3pt
\global\nulldelimiterspace = 0pt



\def\endignore{}
\def\ignore #1\endignore{}

\newcount\dflag
\dflag = 0


\def\monthname{\ifcase\month
\or Jan\ \or Feb\ \or Mar\ \or Apr\ \or May\ \or June\ %
\or July\ \or Aug\ \or Sept\ \or Oct\ \or Nov\ \or Dec\ 
\fi}




\def\endid{}
\def\id#1\endid{\number\day\ \monthname \number\year
\hfill #1}

\def\endtitle{}
\def\title#1\endtitle{\vskip.15in\titlefont
\global\baselineskip = 2\baselineskip
#1\vskip.3in
\baselineskip = 0.5\baselineskip\sectionfont}

\def\lblfoot{This work was supported by the Director, Office of Energy
Research, Office of High Energy and Nuclear Physics, Division of High
Energy Physics of the U.S. Department of Energy under Contract
DE-AC03-76SF00098.}

\def\endauthors{}
\def\authors#1\endauthors{
#1\if\dflag = 0
\footnote{}{\noindent\lblfoot}\fi}

\def\endabstract{}
\def\abstract#1\endabstract{\vskip .2in%
\centerline{\sectionfont\bf Abstract}%
\vskip .1in%
\noindent#1%
\ifnum\dflag = 0
\footline = {\hfil}\pageno = 0
\vfill\eject
\pageno = 1\footline{\centerline{\sectionfont\folio}}
\fi\ifnum\dflag = 2
\footline = {\hfil}\pageno = 0
\fi}

\newcount\nsection
\newcount\nsubsection

\def\section#1{\global\advance\nsection by 1
\global\nsubsection = 0
\bigskip\noindent
\centerline{\sectionfont\bf\number\nsection.\ #1}
\medskip\sectionfont\par\nobreak}

\def\subsection#1{\global\advance\nsubsection by 1
\bigskip\noindent
\centerline{\sectionfont \it \number\nsection.\number\nsubsection.\ #1}
\medskip\rm\par\nobreak}

\def\appendix#1#2{\bigskip\noindent%
\centerline{\sectionfont \bf Appendix #1.\ #2}
\medskip\rm\par\nobreak}


\newcount\nref
\global\nref = 1

\def\ref#1#2{\xdef #1{[\number\nref]}
#1
\ifnum\nref = 1\global\xdef\therefs{\noindent[\number\nref] #2\ }
\else
\global\xdef\oldrefs{\therefs}
\global\xdef\therefs{\oldrefs\vskip.1in\noindent[\number\nref] #2\ }%
\fi%
\global\advance\nref by 1
}

\def\listrefs{\vfill\eject\section{References}\therefs}


\newcount\cflag
\newcount\nequation
\global\nequation = 1
\def\eqlabel{(1)}

\def\nexteqno{\ifnum\cflag = 0
\global\advance\nequation by 1
\fi
\global\cflag = 0
\xdef\eqlabel{(\number\nequation)}}

\def\lasteqno{\global\advance\nequation by -1
\xdef\eqlabel{(\number\nequation)}}

\def\label#1{\xdef #1{(\number\nequation)}
\ifnum\dflag = 1
{\escapechar = -1
\xdef\draftname{\teenyfont\string#1}}
\fi}

\def\clabel#1#2{\xdef\eqlabel{(\number\nequation #2)}
\global\cflag = 1
\xdef #1{\eqlabel}
\ifnum\dflag = 1
{\escapechar = -1
\xdef\draftname{\string#1}}
\fi}

\def\cclabel#1#2{\xdef\eqlabel{#2)}
\global\cflag = 1
\xdef #1{\eqlabel}
\ifnum\dflag = 1
{\escapechar = -1
\xdef\draftname{\string#1}}
\fi}


\def\eeq{}

\def\eqnn #1\eeq{$$ #1 $$}

\def\eq #1\eeq{\xdef\draftname{\ }
$$ #1
\eqno{\eqlabel \rlap{\ \draftname}} $$
\nexteqno}



\def\eol{& \eqlabel \rlap{\ \draftname} \crcr
\nexteqno
\xdef\draftname{\ }}

\def\eeol{& \eqlabel \rlap{\ \draftname}
\nexteqno
\xdef\draftname{\ }}

\def\eolnn{\cr
\global\cflag = 0
\xdef\draftname{\ }}


\def\eqa #1\eeq{\xdef\draftname{\ }
$$ \eqalignno{ #1 } $$
\global\cflag = 0}


\newcount\nfig
\global\nfig = 1

\def\fg#1\efig{\vskip .5in\noindent Fig.\ \number\nfig:\ #1%
\global\advance\nfig by 1}



\def\eg{{\it e.g.\/}}

\def\etc{{\it etc}}



\def\jref#1#2#3#4{{\it #1} {\bf #2}, #3 (#4)}

\def\NPB#1#2#3{\jref{Nucl.\ Phys.}{B#1}{#2}{#3}}
\def\PA#1#2#3{\jref{Physica}{#1A}{#2}{#3}}
\def\PLB#1#2#3{\jref{Phys.\ Lett.}{#1B}{#2}{#3}}

\def\PRD#1#2#3{\jref{Phys.\ Rev.}{D#1}{#2}{#3}}

\def\PRL#1#2#3{\jref{Phys.\ Rev.\ Lett.}{#1}{#2}{#3}}


\def\to{\mathop{\rightarrow}}
\def\too{\mathop{\longrightarrow}}


\def\myint{\int\mkern-5mu}
\def\frac#1#2{{{#1} \over {#2}}\,}  


\def\Dsl{\hbox{/\kern-.6000em\rm D}} 



\def\scr#1{{\cal #1}}
\def\op#1{{\widehat #1}}
\def\mybar#1{\kern 0.8pt\overline{\kern -0.8pt#1\kern -0.8pt}\kern 0.8pt}
\def\sla#1{\raise.15ex\hbox{$/$}\kern-.57em #1}
\def\Sla#1{\kern.15em\raise.15ex\hbox{$/$}\kern-.72em #1}

\def\roughly#1{\mathrel{\raise.3ex\hbox{$#1$\kern-.75em%
    \lower1ex\hbox{$\sim$}}}}
\def\lsim{\roughly<}


\def\tr{\mathop{\rm tr}}


\def\bra#1{\langle #1 |}
\def\ket#1{| #1 \rangle}
\def\braket#1#2{\langle #1 | #2 \rangle}



\hyphenation{ba-ry-on ba-ry-ons ano-ma-ly ano-ma-lies}

\def\al{\alpha}
\def\del{\delta}
\def\Del{\Delta}
\def\gam{\gamma}

\def\ep{\epsilon}

\def\Lam{\Lambda}

\def\Om{\Omega}
\def\sig{\sigma}
\def\Sig{\Sigma}

\def\ChPT{\raise.45ex\hbox{$\chi$}PT}

\def\rhs{right-hand side}
\def\lhs{left-hand side}


\def\MeV{{\rm \ MeV}}

\hyphenation{ba-ry-on ba-ry-ons}


\def\bb{\bra{\scr B_0}}
\def\bk{\ket{\scr B_0}}

\def\rbra#1{(#1|}
\def\rket#1{|#1)}
\def\rbraket#1#2{(#1|#2)}

\def\op#1{\mathord{\{#1\}}}

\def\point#1{{\sl (}{\it #1}{\sl )}}

\def\up{\uparrow}
\def\down{\downarrow}


\id
LBL-35539, NSF-ITP-94-42
\endid
\rightline{hep-ph/9405271}

\title
\vskip-0.05in
\centerline{Baryons with Many Colors and Flavors}
\endtitle

\authors
\vskip-.05in
\centerline{Markus A. Luty}
\vskip .1in
\centerline{\it Lawrence Berkeley Laboratory}
\centerline{\it 1 Cyclotron Road}
\centerline{\it Berkeley, California 94720$\,^*$\footnote{}
{\hskip-.27in ${}^*$ \rm Permanent address}}
\vskip .1in
\centerline{and}
\vskip .1in
\centerline{\it Institute for Theoretical Physics}
\centerline{\it University of California}
\centerline{\it Santa Barbara, California 93106}
\footnote{}{\hskip-.27in
This research was supported in part by the National Science Foundation under
Grant PHY89-04035, and by the Department of Energy under Contract
DE-AC03-76SF00098.}
\endauthors

\abstract
Using recently-developed diagrammatic techniques, I derive some general
results concerning baryons in the $1/N$ expansion, where $N$ is the number
of QCD colors.
I show that the spin-flavor relations which hold for baryons in the
large-$N$ limit, as well as the form of the corrections to these relations
at higher orders in $1/N$, hold even if $N_F / N \sim 1$, where $N_F$ is
the number of light quark flavors.
I also show that the amplitude for a baryon to emit $n$ mesons is
$O(1 / N^{n / 2 - 1})$, and that meson loops attached to baryon lines are
unsupressed in the large-$N$ limit, independent of $N_F$.
For $N_F > 2$, there are ambiguities in the extrapolation away from $N = 3$
because the baryon flavor multiplets for a given spin grow with $N$.
I argue that the $1/N$ expansion is valid for baryons with spin $O(1)$
and {\it arbitrary} flavor quantum numbers, including \eg\ baryons with
isospin and/or strangeness $O(N)$.
This allows the formulation of a large-$N$ expansion in which it is not
necessary to identify the physical baryons with particular large-$N$
states.
$SU(N_F)$ symmetry can be made manifest to all orders in $1/N$, yet group
theory factors must be evaluated explicitly only for $N_F = N = 3$.
To illustrate this expansion, I consider the non-singlet axial currents,
baryon mass splittings, and matrix elements of $\mybar ss$ and
$\mybar s \gam_\mu \gam_5 s$ in the nucleon.
\endabstract


\section{Introduction}

A great deal has been learned about QCD by studying the large-$N$ limit,
where $N$ is the number of colors
\ref\tHooft{G. 't Hooft, \NPB{72}{461}{1974}; \NPB{75}{461}{1974};
for a review, see S. Coleman, {\it Aspects of Symmetry},
(Cambridge, 1985).}.
The large-$N$ limit for baryons is more subtle than for mesons because
baryons contain $N$ quarks, and the structure that emerges is much richer
\ref\Witten{E. Witten, \NPB{160}{57}{1979}.}.
It has been known for some time that in the large-$N$ limit, static baryon
matrix elements obey the nonrelativistic quark-model relations
\ref\Skyrme{K. Bardakci, \NPB{243}{197}{1984};
A. V. Manohar, \NPB{248}{19}{1984}.}
\ref\GS{J.--L. Gervais and B. Sakita, \PRL{52}{87}{1984};
\PRD{30}{1795}{1984}.}.
There has been a recent revival of interest in this problem is due to the
results of
refs.\ \ref\UCSD{R. Dashen and A. V. Manohar, \PLB{315}{425}{1993};
\PLB{315}{438}{1993}; E. Jenkins, \PLB{315}{441}{1993}.},
which showed (for example) that many interesting large-$N$ relations
have corrections starting at $O(1/N^2)$.
These results have been extended using a variety of different methods
\ref\bigUCSD{R. Dashen, E. Jenkins, and A. V. Manohar, UCSD/PTH 93-21,
hep-ph/9310379.}
\ref\Harvard{C. Carone, H. Georgi, and S. Osofsky, \PLB{322}{227}{1994}.}
\ref\us{M. A. Luty and J. March--Russell, LBL-34778, hep-ph/9310369,
submitted to {\it Nucl.\ Phys.} {\bf B}.}.
In this paper, I will use the formalism of ref.\ \us\ to derive some general
results concerning the $1/N$ expansion for baryons:

$\bullet$ I review the argument of ref.\ \us\ and point out that the main
conclusions remain valid when $N_F \sim N$, where $N_F$ is the number of
light quark flavors.
Specifically, the spin-flavor relations which hold in the large-$N$ limit,
as well as the form of the corrections at higher orders in $1/N$ are
independent of $N_F$.
This is in contrast to many large-$N$ results for mesons (such as Zwieg's
rule), which only hold if $N_F \ll N$.

$\bullet$ I derive some general results concerning baryon--meson
interactions using diagrammatic arguments.
I show that the amplitude for a baryon to emit $n$ mesons is
$\lsim O(1/N^{n / 2 - 1})$, independent of $N_F$.
(This result was obtained in ref.\ \Witten\ using a Hartree--Fock
description of baryons made of heavy quarks.)
I also show that meson loops attached to baryon lines in
an effective hadronic field theory are not suppressed by powers of $1/N$,
again independent of $N_F$.
This agrees with explicit calculations in chiral perturbation theory
\UCSD\bigUCSD\us.

$\bullet$ Next, I consider the $1/N$ expansion for $N_F > 2$.
In this case, the baryon $SU(N_F)$ representations grow with $N$, and there
are ambiguities in how to extrapolate the physical baryons to $N > 3$.
I argue that baryons with spin $O(1)$ and {\it arbitrary} flavor quantum
numbers have a well-defined large-$N$ limit, including \eg\ states with
isospin and/or strangeness $O(N)$.
The methods of ref.\ \us\ can then be used to give a $1/N$ expansion with
manifest $SU(N_F)$ symmetry which describes {\it all} the lowest-lying
baryons with spin $O(1)$.
In this expansion, it is not necessary to identify the physical baryons with
particular large-$N$ states.
In addition, the expansion is calculationally very simple, since explicit
calculations of matrix elements need only be carried out for $N = N_F = 3$.
I briefly discuss the relation between this expansion and the one proposed
in ref.\ \bigUCSD.

$\bullet$ I give several examples to show how the expansion described above
works in practice.
I consider matrix elements of non-singlet axial currents, baryon mass
splittings due to quark current masses, and matrix elements of
$\mybar s \gam_\mu \gam_5 s$ and $\mybar ss$ in nucleon states.
More detailed applications which include both $SU(N_F)$ breaking and $1/N$
corrections will be presented elsewhere (see \eg\ ref.\
\ref\usagain{M. A. Luty, J. March--Russell, and M. White, LBL-35598,
CfPA-94-TH-25, hep-ph/9405272.}).

\section{Baryons for large $N$ and $N_F$}

In this section, I briefly review the argument of ref.\ \us\ and show that
the main conclusions continue to hold even if $N_F \sim N$.\footnote{$^*$}
{I do not consider the case $N_F \gg N$, since asymptotic freedom is lost in
this limit, and there is presumably no confinement or chiral symmetry
breaking.}
One might expect that this large-$N$ limit to be the most phenomenologically
relevant in the real world, where $N = N_F = 3$.
There is some evidence that this is so: the $\eta'$ can be regarded as a
$U(1)_A$ Nambu--Goldstone boson in the large-$N$ limit with $N_F \ll N$
\ref\CSB{E. Witten, \NPB{156}{269}{1979};
P. Di Vecchia, \PLB{85}{357}{1979};
G. Veneziano, \NPB{159}{213}{1979};
S. Coleman and E. Witten, \PRL{45}{100}{1980}.},
but the explicit breaking due to the anomaly are $O(N_F / N)$, so this
picture breaks down if $N_F \sim N$.
In the real world, the the $\eta'$ does not look much like a Nambu--Goldstone
boson: for example, $m_{\eta'} = 958 \MeV$.

I begin by discussing the baryon masses in the limit of exact $SU(N_F)$
flavor symmetry.
The idea is to write a {\it perturbative} expansion for the baryon masses at
the quark level and use this to derive the $N$-dependence of physical
quantities, working to all orders in the QCD coupling constant.
It is not clear whether the perturbative expansion converges in any
meaningful sense, but I will assume that the $N$-dependence of the
perturbative expansion is reliable.
This type of reasoning is commonly used for the meson sector \Witten\ and
is known to work in $1+1$ dimensions
\ref\twod{G. 't Hooft, \NPB{75}{461}{1974};
C. G. Callan, N. Coote, and D. J. Gross, \PRD{13}{1649}{1976};
M. B. Einhorn, \PRD{14}{3451}{1976}.}.

The perturbative expansion for the baryon mass closely parallels the
perturbative expansion for the vacuum energy.
The starting point for the expansion of the baryon mass is the state $\bk$,
which is an $N$-quark state with the same quantum numbers as a physical
1-baryon state.
The state $\bk$ plays the same role as the ``no-particle'' state $\ket 0$ in
the evaluation of the vacuum energy; it is defined by
\eq
\label\thestate
\bk \equiv \scr B^{a_1 \al_1 \cdots a_N \al_N}
\epsilon^{A_1 \cdots A_N}
a^\dagger_{a_1 \al_1 A_1} \cdots
a^\dagger_{a_N \al_N A_N} \ket{0},
\eeq
where $a_1, \ldots, a_N = 1, \ldots, N_F$ are flavor indices,
$\al_1, \ldots, \al_N = \,\uparrow, \downarrow$ are spin indices,
and $A_1, \ldots, A_N = 1, \ldots, N$ are color indices.
The operators $a^\dagger$ in eq.\ \thestate\ create a quark with
definite flavor, spin, and color, in a perturbative 1-quark state
$\ket\phi$.
The details of the state $\ket\phi$ are irrelevant for the argument.
The tensor $\scr B$ specifies the spin and flavor quantum numbers of the
state $\bk$.

The diagrammatic expansion for the baryon mass is obtained as follows:
\eqa
\bra{\scr B_0} e^{-iH t} \bk &= \sum {\rm diagrams} \eolnn
\label\limit
&\too |\braket{\scr B_0}{\scr B}|^2 e^{-iE_{\scr B} t}
\quad{\rm as}\quad t \to -i\infty, \eeol
\eeq
where $H$ is the hamiltonian for the system; $\ket{\scr B}$ is the
lowest-energy 1-baryon state with the same quantum numbers as
$\ket{\scr B_0}$, and $E_{\scr B}$ is the energy of the state $\ket{\scr B}$.
In order to match the $t$-dependence of the \rhs\ of eq.\ \limit, the diagrams
must exponentiate, so that
\eq
\label\condiag
-iE_{\scr B}t = \sum {\rm connected\ diagrams}.
\eeq
Because the state $\ket{\scr B_0}$ contains quarks, the diagrams are somewhat
more complicated than the ones appearing in the perturbative expansion of the
vacuum energy.
Typical diagrams are shown in fig.\ 1.
Each diagram corresponds to a matrix element in the state $\bk$ of an operator
containing an equal number of creation and annihilation operators for the
1-quark state $\ket\phi$.

This discussion has been very sketchy, since I will only need topological
properties of the diagrams in this paper.
More details can be found in ref.\ \us.

Subtracting away the vacuum energy, one obtains an expression for the
baryon mass
\eq
\label\themass
M_{\scr B} = \sum_r c_r \bb {\scr O}^{(r)} \bk,
\eeq
where ${\scr O}^{(r)}$ is an $r$-body operator; that is, it contains $r$
creation and $r$ annihilation operators.
It must be a singlet under $SU(N_F)$ as well as $SU(2)_{\rm rot}$ spatial
rotations in the $\scr B$ rest frame.

To determine the $N$-dependence of the coefficients $c_r$ in eq.\ \themass,
one can look at typical graphs (such as those in fig.\ 1) to conclude that
\eq
\label\ncount
c_r \lsim \frac 1{N^{r - 1}}.
\eeq
{\it This result holds even if $N_F \sim N$\/}.
The reason is that each quark--gluon vertex is suppressed by $1/\sqrt N$, and
at least $r - 1$ gluon lines are required to form a connected $r$-body
operator.
If $N_F \sim N$, quark loops are not suppressed; each quark loop gives a
factor $O(N_F / N) \sim 1$ due to the sum over quark flavors, but
eq.\ \ncount\ still holds.
This is the main result of this section.

A similar result holds for baryon matrix elements:
\eq
\bra{\scr B'} T \scr O_1(p_1) \cdots \scr O_n(p_n) \ket{\scr B} =
\sum_{\ell, r} F_{\ell, r}(p_1, \ldots, p_n)
\bra{\scr B'_0} \scr O^{(r)}_\ell \ket{\scr B_0},
\eeq
where
\eq
\label\ncounttoo
F_{\ell, r} \lsim \frac 1{N^{r - 1}},
\eeq
{\it even if} $N_F \sim N$.
Here, $F_{\ell, r}$ is a ``form factor'' depending on the kinematics and
transforming under some irreducible representation (labelled by $\ell$) of
$SU(N_F) \times SU(2)_{\rm rot}$.\footnote{$^*$}
{Eq.\ \ncounttoo\ must be modified if the representation $\ell$ is a
$SU(N_F)$ singlet.
See ref.\ \us, and section 5.3 below.}
$\scr O^{(r)}_\ell$ is an $r$-body operator which also transforms according
to an irreducible representation of $SU(N_F) \times SU(2)_{\rm rot}$.

In ref.\ \us, it is shown that eqs.\ \ncount\ and \ncounttoo\ give rise to
the nonrelativistic quark-model relations for baryon masses and matrix
elements in the large-$N$ limit, as well as a classification of corrections
to all orders in $1/N$.
The new feature of the above discussion is the observation that these
results hold even if $N_F \sim N$.

\section{Meson--baryon interactions in the $1/N$ expansion}

I now consider the question of meson--baryon interactions in the $1/N$
expansion.
I will use an argument similar to the one used by Witten \Witten\ to
determine the $N$-dependence of {\it meson} amplitudes for large $N$.
I briefly review these arguments here, and show how they can be extended
to processes involving baryons.

To study meson masses, consider the 2-point function
$\bra\Om T J(x) J(0) \ket\Om$, where $\ket\Om$ is the (exact)
vacuum state and $J$ is a quark bilinear operator with the right quantum
numbers to create a 1-meson state.
In the large-$N$ limit with $N_F \ll N$, the leading graphs contributing to
the 2-point function are planar graphs containing no internal quark loops,
as illustrated in fig.\ 2.
Witten's observation is that a cut of such a diagram contains only one
color-singlet combination of quark fields $q$ and gluon fields $G$ with
color indices contracted as
$\sim \mybar q_{A_1} {G^{A_1}}_{A_2} \cdots {G^{A_{n - 1}}}_{A_n} q^{A_n}$.
The intermediate state is therefore expected to correspond to a 1-meson
state due to confinement.
Therefore, the spectral representation for the 2-point function can be
written
\eq
\label\spectral
\bra\Om T J(x) J(0) \ket\Om =
\sum_\sig \myint \frac{d^4 q}{(2\pi)^4} e^{-iq\cdot x}
\frac{|\bra \sig J(0) \ket\Om|^2}{q^2 - m_\sig^2 + i0+}
+ {\rm subtractions},
\eeq
where the sum is over 1-meson states $\sig$;
the ``subtractions'' are local terms required by the short-distance behavior
of the 2-point function.
Eq.\ \spectral\ has the form of a sum of tree diagrams in a field theory of
mesons; the subtractions correspond to local counterterms.

To study meson interactions, consider the $n$-point function
$\bra\Om T J(x_1) \cdots J(x_n) \ket\Om$.
By considering the possible cuts of the quark diagrams contributing to the
$n$-point functions in the large-$N$ limit, one finds that the only
possible intermediate states are 1-meson states (see fig.\ 3).
Since the resulting amplitudes obey unitarity and crossing, it is not hard
to see that they are equivalent to the tree approximation of a field theory
of mesons.

Subleading contributions in $1/N$ include diagrams involving internal quark
loops, as well as non-planar diagrams.
It can be shown that diagrams suppressed compared to the leading diagrams by
$1/N^L$ can be cut into at most $L + 1$ color-singlet combinations of
quark and gluon fields, and therefore the spectral representation can
contain intermediate states with at most $L + 1$ color-singlet particles
\Witten.
An example at order $1/N$ is shown in fig.\ 4.
Again, these amplitudes must satisfy crossing and unitarity, and it is
extremely plausible that they can be represented as the $L$-loop
approximation to a hadronic lagrangian.
(This conclusion would follow from Weinberg's ``theorem'' that any
consistent set of amplitudes arises from some quantum field theory
\ref\Weinberg{S. Weinberg, \PA{96}{327}{1979}.}.)

Note that if $N_F \sim N$, quark loops are no longer suppressed because of
the sum over quark flavors, and hence meson interactions are no longer
described by the tree approximation of a lagrangian in the large-$N$ limit.
However, there are other properties of meson interactions for large $N$
which continue to hold even if $N_F \sim N$.
For example, note that
\eq
\label\fscale
f_\sig \equiv \bra \sig J \ket\Om \sim \sqrt{N},
\eeq
since the leading contributions to the \rhs\ of eq.\ \spectral\ are $O(N)$
(even if $N_F \sim N$).
Also, the amplitudes for meson scattering are related to the $n$-point
functions discussed above by LSZ reduction:
\eq
\scr A(\sig_1 \cdots \sig_k \to \sig_{k + 1} \cdots \sig_n)
\sim \frac 1{f_\sig^n} \bra\Om T J \cdots J \ket\Om
\sim \frac 1{N^{n/2 - 1}},
\eeq
since the $n$-point function is $O(N)$ (even if $N_F \sim N$).

To study meson--baryon interactions, I again apply the principle that
physical hadrons in intermediate states correspond to color-singlet
combinations of quark and gluon fields.
When counting color singlets in diagrams involving baryons, one must take
into account the fact that the color lines which attach to the
baryon state combine with the other color indices in $\bk$ to form a color
singlet.
With this in mind, it is easy to count color singlets in the double line
notation.
For example, the diagram in fig.\ 5 is a contribution to the baryons mass
which can give rise to an intermediate state containing a meson and a
baryon.
Evaluating fig.\ 5 gives
\eq
{\rm fig.\ 5} \sim
\frac 1N \bra{\scr B_0} \,\hbox{2-body operator}\, \ket{\scr B_0}
\lsim N.
\eeq
(Note that the form of this contribution agrees with the general result
of eqs.\ \themass\ and \ncount.)
It is not hard to see that one can draw diagrams which give $O(N)$
contributions to the baryons mass, and which have arbitrary numbers of
intermediate color-singlet states by ``iterating'' fig.\ 5 (see fig.\ 6).
Therefore, in a hadronic theory of baryons and mesons, loops of meson
lines attached to baryons are not suppressed by powers of $1/N$, even if
$N_F \ll N$;
the $N$ dependence of loop contributions is constrained only by
eq.\ \themass.
This discussion has focussed on baryon masses, but similar results can
easily be seen to hold for matrix elements as well.
This agrees with explicit calculations in baryon chiral perturbation theory
for large $N$ \UCSD\bigUCSD\us.

To determine the $N$-dependence of the amplitude for a baryon to emit $n$
mesons, consider the $n$-point function
$\bra{\scr B'} T J(x_1) \cdots J(x_n) \ket{\scr B}$.
In the diagrammatic expansion, this matrix element is given as a sum of
connected diagrams with $n$ insertions of the operator $J$;
it is easy to see that it is $\lsim N$ in the large-$N$ limit.
(An example of a diagram which contributes to meson--baryon scattering is shown
in fig.\ 7.)
On the other hand, the matrix element is related to the desired amplitude via
LSZ reduction:
\eq
\label\namp
\scr A(\scr B \sig_1 \cdots \sig_k \to \scr B' \sig_{k + 1} \cdots \sig_n)
\sim \frac 1{f_\sig^n} \bra{\scr B'} T J \cdots J \ket{\scr B}
\lsim \frac 1{N^{n/2 - 1}}.
\eeq
This result was derived ref.\ \Witten\ using a Hartree--Fock picture.
However, it is worth noting that the arguments of ref.\ \Witten\ can be
made precise only for heavy quarks in $3 + 1$ dimensions, and that the
argument given here extends the result to the case $N_F \sim 1$.

These results may appear paradoxical, since the amplitude to emit a single
meson is $O(\sqrt N)$, so one might expect the amplitude to emit $n$
mesons to be $O(N^{n / 2})$.
At the quark level, this expectation can be seen to be false because because
the diagrams which contribute to the physical amplitude must be connected,
and therefore the $N$ dependence arising from combinatoric factors in
diagrams with extra operator insertions does not factor.
In an effective field theory of mesons and baryons, the only $N$ dependence
is in the coupling constants, and the breakdown of factorization occurs
because of cancellations among different graphs \UCSD\bigUCSD\us.
(This has been checked explicitly only for small values of $n$.)
Similar mechanisms resolve the apparent contradiction between
eq.\ \namp\ and the fact that loop contributions are unsupressed.
This a nice example of the duality between the quark-level and hadronic
descriptions of baryons for large $N$.

\section{Baryons with large quantum numbers}

For $N_F = 2$ and $N \gg 1$, the physical baryon states have quantum numbers
$I = J = \frac 12, \frac 32, \ldots$, and it is clear that one should identify
the physical baryons with the large-$N$ states with the same isospin and spin
quantum numbers.
For $N_F > 2$, the size of the $SU(N_F)$ representations associated with a
given spin $J$ grows with $N$ (see fig.\ 8);
for example, for $N_F = 3$, there are baryons with spin $J$ and strangeness
$S = 0, -1, \ldots, -(\frac 12 N + J)$.
It is therefore not obvious how to extrapolate the physical baryon states
to $N > 3$.
For example, the $\Xi$ baryons have maximal strangeness for $N = 3$, and one
could reasonably extrapolate them to either $S = -2$ or
$S = -\frac 12(N + 1)$.

In this section, I argue that the expansion can be formulated so that there
is no need to identify the physical baryons with particular large-$N$
states.
The key point is the observation that the $1/N$ expansion holds for baryons
with spin $J \sim 1$ and {\it all possible flavor quantum numbers},
including \eg\ $S \sim -N$.

I now use the arguments of the previous section to show that the $1/N$
expansion for meson--baryon scattering amplitudes can break down only if the
amplitude is sensitive to the physical process of a baryon decaying to a
single meson.\footnote{$^*$}
{In this paper, I consider only processes involving momentum transfers which
are $O(1)$ in the large-$N$ limit.
If the momentum transfer grows with $N$, there is additional $N$ dependence
in the quark-level diagrams arising from the kinematics.
For example, quarks with momenta $O(N)$ have vanishing overlap with the
states $\ket{\scr B_0}$ in the large-$N$ limit.
Effects such as these tend to suppress exclusive processes
involving momentum transfer $O(N)$ by factors of $e^{-N}$ \Witten.}
For example, the width for a baryon of spin $J \sim N$ to decay to a baryon
of spin $J - 1$ plus a meson is $O(N)$, since the amplitude is $O(\sqrt N)$
and all kinematic invariants are $O(1)$.
(The mass difference between the baryons is at least $O(1)$ in this case
\us.)
Therefore, baryons with spin $J \sim N$ do not have a well-defined large-$N$
limit.
On the other hand, the width for a baryon of spin $J \sim 1$ to decay to a
baryon of spin $J - 1$ plus a pion is $O(1/N^2)$ (in the chiral limit),
because the mass difference between the baryons is $O(1/N)$, reducing the
phase space for the decay.
Baryons with spin $J \sim 1$ are therefore expected to be narrow states in
the large-$N$ limit.

Applying these criteria, one can see that the $1/N$ expansion can be applied
to baryons with $J \sim 1$ and arbitrary quantum numbers.
For example, one can consider decays of a baryon with strangeness $S$ to a
baryon with strangeness $S + 1$ plus a meson with strangeness $-1$.
If the initial and final baryons are in the same $SU(N_F)$ multiplet, then
the decay will be kinematically forbidden near the chiral limit, since the
mass difference between the baryons is $O(m_s)$ (see below) while the
lightest mesons with strangeness $-1$ are kaons with mass $O(m_s^{1/2})$.
If the initial and final baryons differ in spin by $O(1)$, the width for
this decay is $O(1/N^2)$ (in the chiral limit).
One can generalize these considerations to processes involving more mesons
to show that baryons with arbitrary strangeness are well-defined narrow
states for large $N$.
Similar arguments can be given for other quantum numbers.
Thus, scattering amplitudes for {\it all} baryons with $J \sim 1$ have a
well-defined $1/N$ expansion.

Similar arguments also hold for matrix elements.
These can grow as powers of $N$, but this simply means that they must be
rescaled to have a finite large-$N$ limit.
(For example, the electric charge operator must be multiplied by $O(1/N)$ in
order that the baryon have charge $O(1)$ in the large-$N$ limit.)

\subsection{The $1/N$ expansion at $N = 3$}

I now show how the $1/N$ expansion can be carried out for physical
quantities at $N = 3$ without ambiguities arising from the extrapolation of
physical baryons to large $N$.
To understand the idea, consider the $1/N$ expansion of a static operator
matrix element:
\eq
\label\opmat
\bra{\scr B'} J(0) \ket{\scr B} = \sum_{\ell, r}
\frac{a_{\ell, r}}{N^{r - 1}}
\bra{\scr B'_0} \scr O^{(r)}_\ell \bk,
\eeq
where the coefficients $a_{\ell, r}$ are order-1 coefficients that depend on
strong interaction dynamics.
The operators $\scr O^{(r)}_\ell$ which appear on the \rhs\ are
$r$-body operators written in terms of the quark creation and annihilation
operators $a^\dagger$ and $a$ for the 1-quark state $\ket\phi$ used to
define $\bk$.
Each of the operators $\scr O_\ell^{(r)}$ on the \rhs\ of eq.\ \opmat\ must
have the same transformation properties under
$SU(N_F) \times SU(2)_{\rm rot}$ as the operator $J(0)$ on the \lhs.

The $1/N$ expansion is carried out in three steps:
\point{i} Classify the possible operators $\scr O_\ell^{(r)}$ that can
appear in eq.\ \opmat.
This classification is explained in ref.\ \us, and I will give some
examples below.
\point{ii} Evaluate the $N$-dependence of the contribution of each operator
$\scr O_\ell^{(r)}$ to determine whether it contributes at the order one
is working.
Of course, the $N$-dependence of an operator depends on the states in which
it is evaluated.
In the expansion considered here, {\it all} of the baryon states with
$J \sim 1$ are included, and so one must keep an operator if its matrix
element in {\it any} pair of states $\ket{\scr B}$ and $\ket{\scr B'}$ is of
sufficiently high order in $N$.
\point{iii} Evaluate the surviving terms on the \rhs\ of eq.\ \opmat\ for
$N = 3$ to obtain the physical results.
Note that $SU(N_F)$ symmetry is manifest at each step, since the operators
$\scr O_\ell^{(r)}$ have well-defined $SU(N_F)$ quantum numbers.

Before giving some examples of the expansion proposed above, I briefly
contrast this expansion to the $1/N$ expansion proposed in ref.\ \bigUCSD.
In ref.\ \bigUCSD\ the baryons at $N = 3$ are extrapolated to baryons with
strangeness $O(1)$ in the large-$N$ limit.
In a world where the physical value of $N$ is very large, this expansion
would work well only for states with small strangeness
($|S| \ll \frac 12 N$).
In the real world, where $N = 3$, it is not clear where to draw the line
between ``large'' and ``small'' strangeness, and it is consistent to assume
that all baryons which arise at $N = 3$ have ``small'' strangeness.
In the language of eq.\ \opmat, this expansion would (in general) differ
from the one proposed above in which operators to keep at a given order in
the $1/N$ expansion, since the expansion of ref.\ \bigUCSD\ requires that
the operators on the \rhs\ be tested only on states with strangeness of
order 1.
The expansion advocated here is therefore more conservative, in the sense
that (in general) more operators are kept at a given order in the expansion.
The ``extra'' terms in this expansion are suppressed by $1/N$ according to
the philosophy of ref.\ \bigUCSD, and experiment can in principle be used
to decide whether these terms are in fact suppressed for $N = 3$.

\section{Examples}

In this section, I give several examples of the expansion described
above.
More detailed phenomenological applications will appear elsewhere
(see \eg\ \usagain).

For specific applications, it is convenient to use the effective
lagrangian described in ref.\ \us, which I briefly review.
Because the baryon mass is of order $N \Lam_{\rm QCD}$, the baryons can be
described using a heavy-particle effective field theory
\ref\heavyB{E. Jenkins and A. V. Manohar, \PLB{255}{558}{1991}.}.
The baryon momentum is written $P = M_0 v + k$, where $M_0 \sim N$ is a
baryon mass and $v$ is a 4-velocity ($v^2 = 1$) which defines the baryon
rest frame.
The effective field theory is then written in terms of baryon fields whose
momentum modes are the residual momenta $k$.
This effective field theory gives an expansion in $1/M_0$ around the
static limit.

For $N$ large, the baryon $SU(N_F)$ representations are large, and it is
convenient to use a Fock-space notation to keep track of baryon quantum
numbers:
the baryons fields are written
\eq
\rket{\scr B(x)} \equiv \scr B^{a_1 \al_1 \cdots a_N \al_N}(x)\,
\al^\dagger_{a_1 \al_1} \cdots \al^\dagger_{a_N \al_N} \rket 0.
\eeq
The $\al^\dagger$'s are {\it bosonic} creation operators which create a
``quark'' with definite flavor and spin, and $\rket 0$ is the Fock
``vacuum'' state;
$a_1, \ldots, a_N$ are $SU(N_F)$ flavor indices and
$\al_1, \ldots, \al_N = \up, \down$ are spin indices in the rest frame
defined by $v$.
\ignore
(In a fully relativistic notation, $\al_1, \ldots, \al_N$ would be Dirac
indices running from 1 to 4, and the fields $\scr B$ would obey the
constraints
$(\sla v)^{\al_j}_\beta \scr B^{a_1 \al_1 \cdots a_j \beta \cdots a_N \al_N} =
\scr B^{a_1 \al_1 \cdots a_N \al_N}$, $j = 1,\ldots, N$.)
\endignore
The baryon kinetic term can then be written
\eq
\scr L_{\rm eff} = \rbra{\scr B} iv^\mu \partial_\mu \rket{\scr B} + \cdots.
\eeq
The couplings of PNGB's to baryons can be introduced using standard
techniques from chiral perturbation theory \us.
(Note that if $N_F \sim N$, the $\eta'$ is not light and is not present in
the effective lagrangian.)

\subsection{Axial Currents}

As a first application, I consider the non-singlet axial currents.
Matrix elements of the axial currents are defined by coupling a source
$\scr A^\mu$ to the axial currents in the QCD lagrangian:
\eq
\del\scr L_{\rm QCD} = \mybar q \scr A^\mu \gam_\mu \gam_5 q,
\qquad q = \pmatrix{u \cr d \cr s \cr},
\eeq
The terms in the effective lagrangian linear in the source can be written
\eq
\del\scr L_{\rm eff} = \rbra{\scr B} \Bigl[
g \op{\scr A^\mu \sig_\mu}
+ \frac hN\, \op{\scr A^\mu} \op{\sig_\mu}
+ \frac{h'}{N^2} \op{\sig^\mu} \op{\scr A^\nu \sig_\nu} \op{\sig_\mu}
+ \cdots \Bigr] \rket{\scr B}.
\eeq
Here $\sig_\mu \equiv (\sla v \gam_\mu - v_\mu) \gam_5$ is the spin
matrix, and ``quark-model'' operators are defined by
\eq
\label\axeff
\op{\scr A^\mu \sig_\mu} \equiv \al^\dagger_{a\al}
{(\scr A^\mu)^a}_b {(\sig_\mu)^\al}_\beta \al^{b\beta},
\eeq
\etc.
The general rule for the $N$-dependence of coefficients in the effective
lagrangian is obtained by imposing eq.\ \ncounttoo.
The result is that $r$-body operators in the effective lagrangian
have coefficients $O(1/N^{r - 1})$.
In particular, $g, h, h' \sim 1$ in the large-$N$ limit.

In order to determine the $N$-dependence of the axial current matrix
elements, one must evaluate the matrix elements of quark-model operators
such as the one in eq.\ \axeff.
In appendix A, it is shown that $\op{\scr A^\mu \sig_\mu}$ has matrix
elements which are $O(N)$ in some baryon states with $J \sim 1$.
(A $r$-body operator can have matrix elements {\it at most} $O(N^r)$.)
This means that $\op{\scr A^\mu \sig_\mu}$ counts as being $O(N)$ in the
expansion described in the previous section.
The operator $\op{\sig^\mu}$ simply measures the total spin of the
baryon, and therefore has matrix elements which are $O(1)$ on all states
with $J \sim 1$.
Therefore, the contributions to axial current matrix elements proportional
to $g$ are $O(N)$, those proportional to $h$ are $O(1)$, and all others are
suppressed by additional powers of $1/N$.

At leading order in $1/N$, the the axial current matrix elements are
determined by a single coupling constant $g$.
The matrix elements of $\op{\scr A^\mu \sig_\mu}$ are exactly what one
would compute in the static quark model, and one obtains the relations
\eq
\label\lorel
D = g, \quad
F = \frac 23 g, \quad
\scr C = -2g,
\eeq
with corrections $O(1/N$).
Here $D$ and $F$ are the usual $SU(3)$ axial couplings, and $\scr C$ is
the decuplet-octet axial coupling defined in
ref.\ \ref\JMdec{E. Jenkins and A. V. Manohar, \PLB{255}{558}{1991}.}.
(In the static quark model, $g = 1$.
In the $1/N$ expansion of QCD, determining $g$ requires a non-trivial
dynamical calculation.)
At subleading order in $1/N$, the axial current matrix elements are
determined by two coupling constants $g$ and $h$.
In appendix A, it is shown that these terms are independent at this
order in the $1/N$ expansion.
For $N_F = 3$,
\eq
\label\corel
D = g, \quad
F = \frac{2 g + h}3, \quad
\scr C = -2g,
\eeq
with corrections $O(1/N^2)$.

The couplings $D$ and $F$ are measured in semileptonic decays, and the
coupling $\scr C$ is measured in decuplet strong decays.
A fit ignoring explicit chiral symmetry breaking gives $D \simeq 0.8$,
$F \simeq 0.5$ and $\scr C \simeq 1.5$
\ref\JMrev{E. Jenkins and A. V. Manohar, in U.--G. Mei\ss ner, {\it ed.},
{\it Effective Field Theories of the Standard Model\/} (World Scientific,
1993).},
so that both the lowest-order
relations in eq.\ \lorel\ and the ``corrected'' relations in
eq.\ \corel\ appear to work well.
(Using the best-fit values for the axial couplings, $h \simeq -0.1$.)
However, chiral symmetry breaking due to the strange quark mass is expected
to be sizable both for the semileptonic decays
\JMdec\ and for the strong decuplet decays
\ref\strong{M. N. Butler, M. J. Savage, and R. P. Springer,
\NPB{399}{69}{1993}.}.
The validity of the $1/N$ expansion including explicit chiral symmetry
breaking will be considered in detail in a future publication.

\subsection{Baryon Mass Differences}

Next, I consider the $SU(N_F)$-breaking baryon mass differences induced by
$m_s \ne 0$.
I assume that $O(m_s)$ and $O(1/N)$ corrections are both
$O(\ep) \sim 30\%$, and expand consistently in powers of $\ep$.

The dependence on $m_s$ occurs through the quark mass spurion
\eq
\label\mspur
m = m_s S, \quad
S = \pmatrix{0 &&&\cr & \ddots &&\cr && 0 &\cr &&& 1 \cr},
\eeq
which transforms as $m \mapsto U m U^\dagger$ under $SU(N_F)$.
The leading terms giving rise to baryon mass splittings are
\eq
\label\lmass
\eqalign{
\del\scr L = & a \rbra{\scr B} \op{m} \rket{\scr B}
+ \frac bN\, \rbra{\scr B} \op{\sig^\mu} \op{\sig_\mu} \rket{\scr B}
+ \frac cN\, \rbra{\scr B} \op{m \sig^\mu}
\op{\sig_\mu} \rket{\scr B} \cr
&\quad + d_0 \rbra{\scr B} \op{m^2} \rket{\scr B}
+ \frac{d_1}N\, \rbra{\scr B} \op{m} \op{m} \rket{\scr B}
+ \frac{d_2}N\, \rbra{\scr B} \op{m \sig^\mu} \op{m \sig_\mu} \rket{\scr B}
+ \cdots. \cr}
\eeq
Using the rules given above, all of the coefficients in this lagrangian are
$O(1)$ in the large-$N$ limit.
The matrix elements of $\op{S}$ and $\op{S \sig^\mu}$ can be $O(N)$ in the
large-$N$ limit, so the maximal $N$-dependence of the matrix elements
relevant for the baryon masses is
\eq
\label\smat
\eqalign{
\op{m} &\sim N m_s \sim N\ep, \cr
\frac 1N\, \op{\sig^\mu} \op{\sig_\mu} &\sim \frac 1N \sim N \ep^2, \cr
\frac 1N\, \op{m \sig^\mu} \op{\sig_\mu} &\sim m_s \sim N\ep^2, \cr
\op{m^2} &\sim N m_s^2 \sim N \ep^2, \cr
\frac 1N\, \op{m} \op{m} &\sim N m_s^2 \sim N \ep^2, \cr
\frac 1N\, \op{m\sig^\mu} \op{m\sig_\mu} &\sim N m_s^2 \sim N \ep^2. \cr}
\eeq
It is easily checked that all other operators are at most $O(N\ep^3)$.

Loops of pseudo Nambu--Goldstone bosons can induce non-analytic dependence
on $m_s$ of the form $m_s^{3/2}$ and $m_s \ln m_s$
\ref\Pagels{L.-F. Li and H. Pagels, \PRL{26}{1204}{1971};
P. Langacker and H. Pagels, \PRD{10}{2904}{1974}.}.
These nonanalytic terms must be multiplied by $r$-body operators with
coefficients $O(1/N^{r - 1})$ in accordance with eqs.\ \themass\ and
\ncount.
The breaking of $SU(N_F)$ in these operators must be proportional to powers
of the spurion $S$ defined in eq.\ \mspur, and it is easy to check that the
operators which can contribute at $O(N m_s^{3/2}) = O(N\ep^{3/2})$ and
$O(N m_s \ln m_s) = O(N\ep \ln\ep)$ are of the same form as those appearing
in eq.\ \lmass.
Therefore, relations which are independent of the coefficients in
eq.\ \lmass\ are not invalidated by the nonanalytic corrections.
The leading corrections to these relations are then
$O(N m_s^{5/2}) = O(N \ep^{5/2})$.

The operators appearing in eq.\ \lmass\ are not all independent:
it is easily checked that
\eq
\op{S^2} = \op{S}, \quad
\op{S \sig^\mu} \op{S \sig_\mu} = -\op{S} \op{S} + 2 \op{S}.
\eeq
Therefore, there are 4 independent operators which determine the 7
independent mass differences of the octet and decuplet isospin multiplets.
This gives 3 independent relations valid to $O(\ep^2)$,\footnote{$^*$}
{Note that mass splittings between states differing in strangeness by 1 unit
induced by the operators in eq.\ \lmass\ are smaller than the estimates in
eq.\ \smat\ by $O(1/N)$.}
which may be written as
\eqa
\label\equal
(M_\Om - M_{\Xi^*}) - (M_{\Xi^*} - M_{\Sig^*}) &=
(M_{\Xi^*} - M_{\Sig^*}) - (M_{\Sig^*} - M_\Del), \eol
\label\deficit
3 M_\Lam + M_\Sig - 2 M_N - 2 M_\Xi &=
(M_{\Sig^*} - M_\Del) - (M_\Om - M_\Xi^*) \eol
\label\newmass
M_{\Xi^*} - M_{\Sig^*} &= M_\Xi - M_\Sig \eeol
\eeq
Eq.\ \equal\ is an ``improved'' version of the decuplet equal-spacing rule,
which holds to $O(m_s^2)$ independently of the $1/N$ expansion
\ref\Jmass{E. Jenkins, \NPB{368}{190}{1992}.}.
Eqs.\ \deficit\ and \newmass\ are non-trivial predictions of the $1/N$
expansion;
eq.\ \deficit\ relates the deficit of the Gell-Mann--Okubo relation to that
of the decuplet equal-spacing rule.
The same relations are derived in ref.\ \bigUCSD\ under rather different
assumptions:
the strangeness of physical baryons was assumed to be $O(1)$ in the
large-$N$ limit, and $m_s$ was {\it not} assumed to be a small parameter.
Under these assumptions, operators proportional to high powers of $\op{S}$
are suppressed by powers of $1/N$, and the mass splittings to $O(1/N^2)$
are determined by the operators in eq.\ \lmass.

\subsection{Strangeness in the Nucleon}

Next, I consider the matrix elements
$\bra p \mybar s \gam_\mu \gam_5 s \ket p$ and $\bra p \mybar ss \ket p$,
which give information on the strange content of the nucleons.
These matrix elements have been considered in the context of the Skyrme
model
\ref\BEK{S. J. Brodsky, J. Ellis, and M. Karliner, \PLB{206}{1988}{309}.}
\ref\DN{J. F. Donoghue and C. R. Nappi, \PLB{168}{105}{1986}.},
and using the Hartree--Fock picture \Harvard.

The leading term in the baryon effective lagrangian which can
give rise to a nonzero value for $\bra p \mybar s \gam_\mu \gam_5 s \ket p$
is
\eq
\label\axterm
\del\scr L = \frac cN \tr(\scr A^\mu)\,
\rbra{\scr B} \op{\sig_\mu} \rket{\scr B},
\eeq
where $c \sim 1$.
The reason for the explicit factor of $1/N$ in the coefficient is that the
leading quark-level diagrams which can give rise to a flavor trace involve a
quark loop (see fig.\ 9).
This factor is there even if $N_F \sim N$, since the $N_F$ dependence is
correctly accounted for by the flavor trace in eq.\ \axterm.
Since $\op{\sig_\mu} = O(1)$ on any baryon states with $J \sim 1$,
$\bra p \mybar s \gam_\mu \gam_5 s \ket p = O(1/N)$.

Matrix elements of $\mybar ss$ can be defined by differentiating with
respect to $m_s$ in the QCD lagrangian.
The leading term which can give rise to a nonzero value for
$\bra p \mybar ss \ket p$ is
\eq
\del\scr L = \frac dN \tr(m)\, \rbra{\scr B} \op{1} \rket{\scr B},
\eeq
where $d \sim 1$.
(This term was not considered above because it does not give rise to baryon
mass differences.)
Since $\op{1} = N$ on any baryon state, $\bra p \mybar ss \ket p = O(1)$.

These results disagree with refs.\ \BEK\DN;
I do not understand the arguments in these papers well enough to
explain this disagreement.

\section{Conclusions}

Using the methods of ref.\ \us, I have derived some general results
concerning the $1/N$ expansion for baryons.
I showed that the form of the $1/N$ expansion is unchanged if $N_F \sim N$
and gave a simple method to carry out a simultaneous expansion in $1/N$
which keeps $SU(N_F)$ symmetry manifest for arbitrary $N$ and $N_F$.
These ideas were illustrated with some simple examples.
Clearly, more work remains to be done.

\section{Acknowledgements}

I thank A. Falk, H. Murayama, M. J. Savage, and especially
J. March--Russell, R. Sundrum, and M. White for many helpful
discussions on the topic of this paper.
I thank the Institute for Theoretical Physics at Santa Barbara for
hospitality while this work was in progress.
This research was supported in part by the National Science Foundation under
Grant PHY89-04035, and by the Director, Office of Energy Research, Office of
High Energy and Nuclear Physics, Division of High Energy Physics of the U.S.
Department of Energy under Contract DE-AC03-76SF00098.

\vfill\eject\appendix{A}{Calculation of matrix elements%
\footnote{$^*$}{}}

\footnote{}{$*$\ The methods used in this appendix
are the result of a collaboration with J. March--Russell.}

The purpose of this appendix is to illustrate the determination of the
$N$ dependence of the matrix elements discussed in the main text.
For this purpose, it is sufficient to consider spin-$\frac 12$ baryons for
$N_F = 3$, although the methods and results discussed here are clearly
more general.

For $N_F = 3$, the most general spin-$\frac 12$ baryon states
can be written
\eq
\label\halfstates
\rket{\scr B} = \scr B^a_{b_1 \cdots b_n} \chi^\al
\al^\dagger_{a\al}
A^{b_1 \kern-0.1em \dagger} \cdots A^{b_n \kern-0.1em \dagger} \rket 0,
\eeq
where
\eq
A^{a \dagger} \equiv \frac 12 \ep^{abc} \ep^{\beta\gamma}
\al^\dagger_{b\beta} \al^\dagger_{c\gamma}
\eeq
creates a pair of quarks in a spin-singlet, flavor $\mybar 3$ state, and
\eq
n \equiv \frac{N - 1}2.
\eeq
Here, $a, b_1, \ldots = u, d, s$ are flavor indices and
$\al, \beta, \ldots = \up, \down$ are spin indices.
Because the $A^\dagger$ operators commute, $\scr B^a_{b_1 \cdots b_n}$ is
totally symmetric in $b_1 \cdots b_n$.

The multiplet defined by eq.\ \halfstates\ contains states which have the
same isospin and strangeness as the octet baryons:
\eqa\label\firststate
\rket{N} &\equiv C_N I^a \chi^\al \al^\dagger_{a\al}
(A^{s\dagger})^n \rket 0, \eol
\rket{\Sig} &\equiv C_\Sig {I^a}_b \chi^\al \al^\dagger_{a\al}
A^{b\dagger} (A^{s\dagger})^{n - 1} \rket 0, \eol
\rket{\Xi} &\equiv C_\Xi I_a \chi^\al \al^\dagger_{s\al}
A^{a\dagger} (A^{s\dagger})^{n - 1} \rket 0, \eol
\label\laststate
\rket{\Lam} &\equiv C_\Lam \chi^\al \al^\dagger_{s\al}
(A^{s\dagger})^n \rket 0, \eeol
\eeq
where $\chi$ ($I$) is the appropriate spin (isospin) tensor.
One can identify these states with the large-$N$ limit of the physical
states, as done in ref.\ \bigUCSD.
However, there are also other natural candidates.
For example, the states
\eq
\label\xiprime
\rket{\Xi'} \equiv C_{\Xi'} I_{a_1 \cdots a_n} \chi^\al
\al^\dagger_{s\al} A^{a_1 \kern-0.1em \dagger} \cdots
A^{a_n \kern-0.1em \dagger} \rket{0}
\eeq
have $I = n/2 = O(N)$, $S = -(n + 1) = O(N)$, the same quantum numbers as
the $\Xi$ baryons for $N = 3$.
The $1/N$ expansion discussed in the main text does not require that the
physical states be identified with specific large-$N$ states, but the states
in eqs.\ \firststate--\laststate\ will be used for illustrative
purposes.

Calculations involving these states can be carried out efficiently using
an occupation number representation.
Define the occupation number states
\eq
\rket{n_1, \ldots, n_6} \equiv (\al_1^\dagger)^{n_1} \cdots
(\al_6^\dagger)^{n_6} \rket 0,
\eeq
using the abbreviation
$u\!\!\up, u\!\!\down, d\!\!\up, d\!\!\down, s\!\!\up, s\!\!\down\ =
1, \ldots, 6$.
Note that the occupation number states defined here are not unit-normalized:
\eq
\rbraket{m_1, \ldots, m_6}{n_1, \ldots, n_6} = n_1! \cdots n_6! \,
\delta_{m_1 n_1} \cdots \delta_{m_6 n_6}.
\eeq
In this notation,
$A^{s\dagger} = \al^\dagger_1 \al^\dagger_4 - \al^\dagger_2 \al^\dagger_3$,
so that
\eqa
\rket{p, \up} &= C_N \al^\dagger_1
(\al^\dagger_1 \al^\dagger_4 - \al^\dagger_2 \al^\dagger_3)^n \rket 0 \eolnn
&= C_N \sum_{k = 0}^n (-1)^k \frac{n!}{k! (n - k)!}
\rket{n - k + 1, k, k, n - k, 0, 0}. \eeol
\eeq
Using this notation, it is easy to compute matrix elements.
\def\ax{\op{\tau_3 \sig_3}}
For example, the operator
\eq
\ax = \al^\dagger_1 \al^{\vphantom\dagger}_1
- \al^\dagger_2 \al^{\vphantom\dagger}_2
- \al^\dagger_3 \al^{\vphantom\dagger}_3
+ \al^\dagger_4 \al^{\vphantom\dagger}_4,
\eeq
has matrix element
\eqa
\rbra{p, \up} \ax \rket{p, \up} &= C_N
\sum_{k = 0}^n (-1)^k \frac{n!}{k! (n - k)!}
\,(n - k + 1 - k - k + n - k)\eolnn
&\qquad\qquad\qquad \times
\rbraket{p, \up}{n - k + 1, k, k, n - k, 0, 0} \eolnn
&= C_N^2 \sum_{k = 0}^n \left[\frac{n!}{k! (n - k)!}\right]^2
(2n - 4k + 1) (n - k + 1)! (k!)^2 (n - k)! \eolnn
&= (n! C_N)^2 \sum_{k = 0}^n (2n - 4k + 1)(n - k + 1) \eeol
\eeq
$C_N$ can be determined by normalizing the state $\rket{p, \up}$ by a
similar calculation.
Combining these results, one obtains
\eq
\rbra{p, \up} \ax \rket{p, \up} = \frac{2n + 3}3.
\eeq
For $N = 3$, the standard quark-model result is recovered;
note that the matrix element is proportional to $N$ for large $N$.
This is because the number of $u$ and $d$ quarks in the protron is of
order $N$, and their contribution to this matrix element adds coherently.

Similar calculations can be used to compute the matrix elements shown in
table 1.
This table substantiates some of the claims made in the main text;
for example, one can see that the operators $\op{\tau_3 \sig_3}$ and
$\op{\tau_3}\op{\tau_3}$ are independent, and that matrix elements of the
operator $\op{S \sig_j} \op{\sig_j}$ are of order the strangeness of the
state.
It is also amusing to compare the matrix elements of the states $\rket{\Xi}$
and $\rket{\Xi'}$ in table 1:
the matrix elements are equal for $N = 3$, but they are very different for
$N \gg 3$.

\vbox{\vskip 20pt \centerline{
\vbox{\offinterlineskip
\halign{\vrule#
&\hfil#\hfil&\vrule#
&\hfil#\hfil&\vrule#
&\hfil#\hfil&\vrule#
&\hfil#\hfil&\vrule#
&\hfil#\hfil&\vrule#
\cr
\noalign{\hrule}
height2pt
&\omit&
&\omit&
&\omit&
&\omit&
&\omit&
\cr
&\quad\quad\quad&
&\quad $\op{\tau_3 \sig_3}$ \quad&
&\quad $\op{\tau_3} \op{\sig_3}$ \quad&
&\quad $\op{S}$ \quad&
&\quad $\op{S \sig_j} \op{\sig_j}$ \quad&
\cr
height2pt
&\omit&
&\omit&
&\omit&
&\omit&
&\omit&
\cr
\noalign{\hrule}
%
%
\noalign{\hrule}
height2pt
&\omit&
&\omit&
&\omit&
&\omit&
&\omit&
\cr
& $p$ &
& $\frac{2n + 3}3$ &
& 1 &
& 0 &
& 0 &
\cr
height2pt
&\omit&
&\omit&
&\omit&
&\omit&
&\omit&
\cr
%
%
\noalign{\hrule}
height2pt
&\omit&
&\omit&
&\omit&
&\omit&
\cr
& $\Lam$ &
& $0\frac{\vphantom 1}{\vphantom 1}$ &
& 0 &
& 1 &
& 3 &
\cr
height2pt
&\omit&
&\omit&
&\omit&
&\omit&
&\omit&
\cr
%
%
\noalign{\hrule}
height2pt
&\omit&
&\omit&
&\omit&
&\omit&
&\omit&
\cr
& $\Sig^+$ &
& $\frac{2n + 2}3$ &
& 3 &
& 1 &
& -1 &
\cr
height2pt
&\omit&
&\omit&
&\omit&
&\omit&
&\omit&
\cr
%
%
\noalign{\hrule}
height2pt
&\omit&
&\omit&
&\omit&
&\omit&
&\omit&
\cr
& $\Xi^0$ &
& $-\frac{2n + 1}9$ &
& 1 &
& 2 &
& 4 &
\cr
height2pt
&\omit&
&\omit&
&\omit&
&\omit&
&\omit&
\cr
%
%
\noalign{\hrule}
height2pt
&\omit&
&\omit&
&\omit&
&\omit&
&\omit&
\cr
& $\Xi'$ &
& $-\frac{n^2}{n + 2}$ &
& $n$ &
& $n + 1$ &
& \quad$\frac{3n^2 + 7n + 6 - 4 \del_{n1}}{n + 2}$\quad &
\cr
height2pt
&\omit&
&\omit&
&\omit&
&\omit&
&\omit&
\cr
\noalign{\hrule}
}}}\vskip 10pt }
{\narrower
Table 1: Assorted ``quark-model'' matrix elements.
The states $p$, $\Lam$, $\Sig^+$, and $\Xi^0$ are defined in
eqs.\ \firststate--\laststate; $\Xi'$ is defined in eq.\ \xiprime, and
has $I_3 = n/2$.
All states have $J_3 = +1/2$.\par}
\vskip 15pt

\ignore
The octet wavefunctions are
\eq
\eqalign{
\rket{N} &= C_N I_N^a \chi^\al \al^\dagger_{a\al} \del^\dagger \rket{0}, \cr
\rket{\Sig} &= C_\Sig I_\Sig^{ab} \chi^\al \ep^{\beta\gam}
\al^\dagger_{a\al} \al^\dagger_{b\beta} \al^\dagger_{s\gamma} \rket{0}, \cr
\rket{\Xi} &= C_\Xi I^a_\Xi \chi^\al \ep^{\beta\gam}
\al^\dagger_{s\al} \al^\dagger_{a\beta} \al^\dagger_{s\gamma} \rket{0}, \cr
\rket{\Lam} &= C_\Lam \chi^\al \al^\dagger_{s\al}
\del^\dagger \rket{0}, \cr}
\eeq
where $a, b, \ldots$ are {\it isospin} indices, the $I$'s are isospin
tensors,
\eq
I_N = \pmatrix{p \cr n \cr}, \quad
I_\Sig = \pmatrix{\Sig^+ & \frac 1{\sqrt 2} \Sig^0 \cr
\frac 1{\sqrt 2} \Sig^0 & \Sig^- \cr}, \quad
I_\Xi = \pmatrix{\Xi^0 \cr \Xi^- \cr},
\eeq
the $\chi$'s are spin tensors, and I have defined
\eq
\del^\dagger \equiv \frac 12 \ep^{ab} \ep^{\al\beta}
\al^\dagger_{a\al} \al^\dagger_{b\beta}.
\eeq
Normalizing these states, gives
\eq
C_N = C_\Sig = C_\Xi = \frac 1{\sqrt 3}, \quad
C_\Lam = -\frac 1{\sqrt 2}.
\eeq
With these phase conventions, the octet baryon matrix is
\eq
B = \pmatrix{\frac 1{\sqrt 2} \Sig^0 + \frac 1{\sqrt 6} \Lam &
-\Sig^+ & p \cr
\Sig^- & -\frac 1{\sqrt 2} \Sig^0 + \frac 1{\sqrt 6} \Lam & n \cr
\Xi^- & -\Xi^0 & \frac 2{\sqrt 6} \Lam \cr}.
\eeq
Note the minus signs in front of $\Sig^+$ and $\Xi^0$.
\endignore


\listrefs
\vfill\eject

\centerline{\bf Figure Captions}
\vskip .2in
\noindent
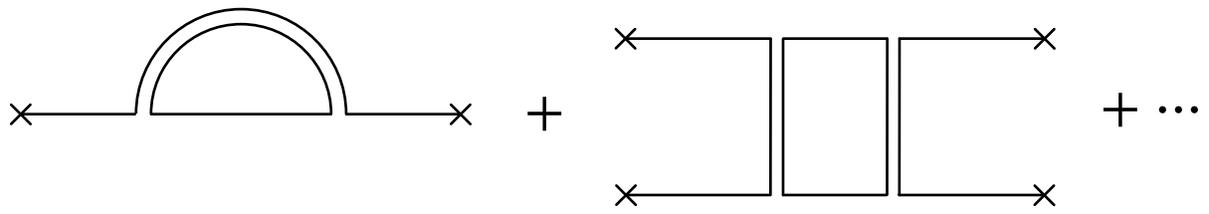
Fig.\ 1.
Typical diagrams contributing to the baryon mass.
Double-line notation is used to keep track of color flow:
single lines denote quarks and double lines denote gluons.
Each diagram corresponds to a term in the expansion of eq.\ \condiag,
consisting of an operator matrix element in the state $\bk$.
The external quark lines ending in a ``$\times$'' denote creation and
annihilation operators $a^\dagger$ and $a$, while the internal lines are given
by the Feynman rules described in ref.\ \us.

\smallskip\noindent
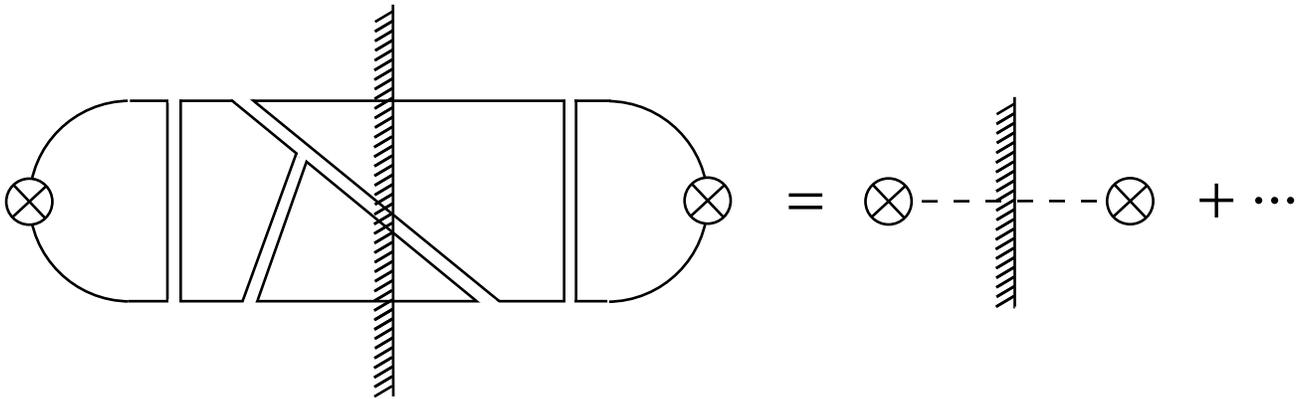
Fig.\ 2.
A typical planar diagram contributing to the 2-point function
$\bra\Om T J J \ket\Om$ at order $N$.
The cuts of this diagram contain only a single color-singlet state, and so
this contribution can only give rise to 1-meson states, as shown on the right.

\smallskip\noindent
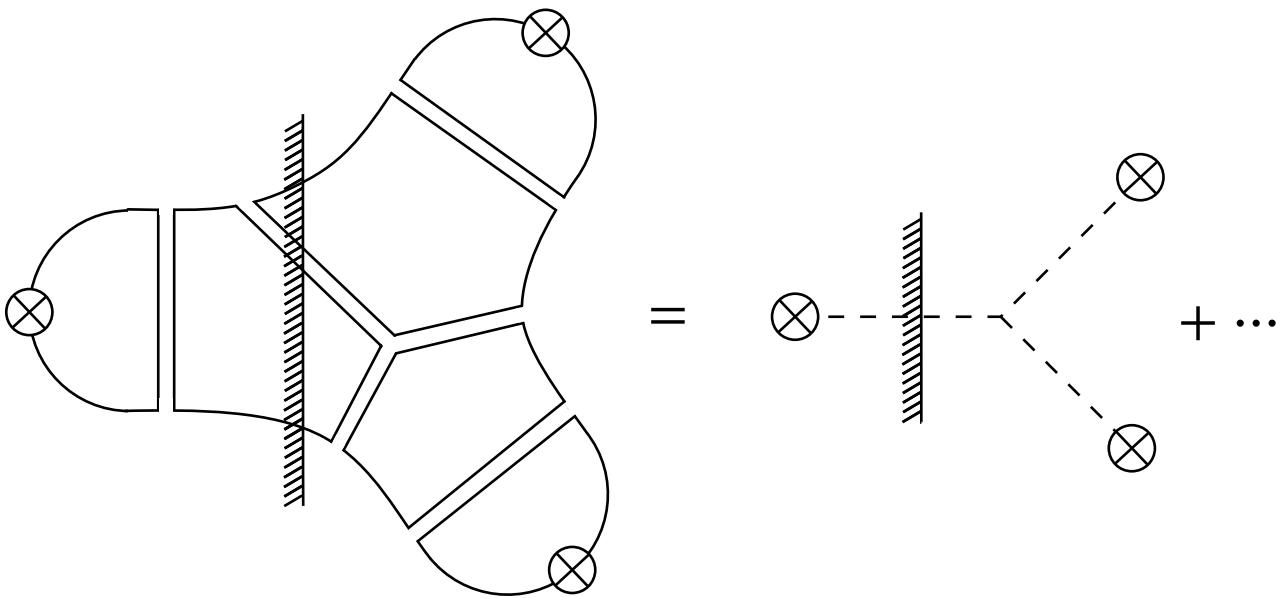
Fig.\ 3.
A typical planar diagram contributing to the 3-point function
$\bra\Om T J J J \ket\Om$ at order $N$.
The cuts of this diagram contains only a single color-singlet state, and so
this contribution can only give rise to tree-level meson processes such the one
shown on the right.

\smallskip\noindent
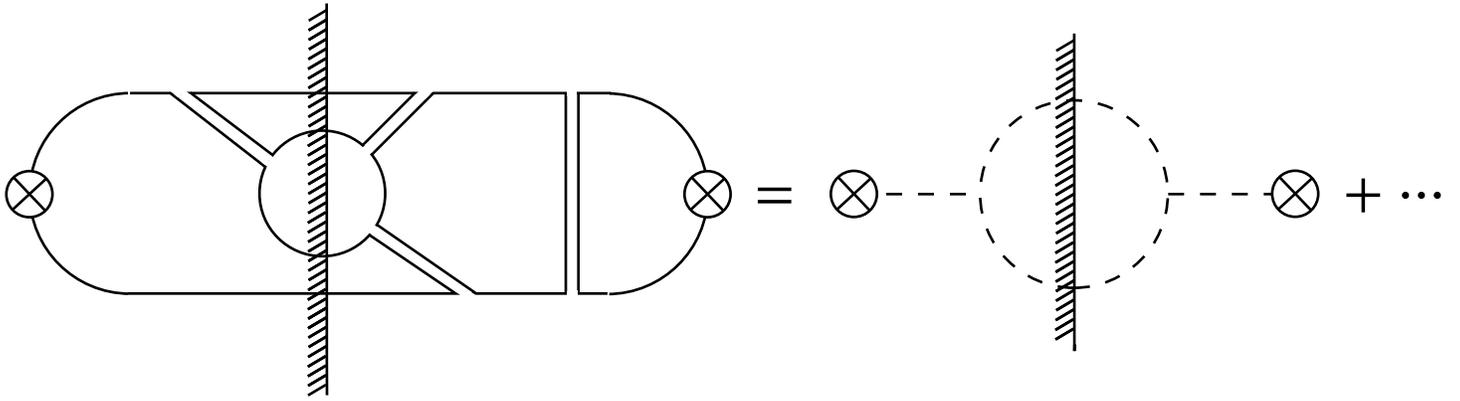
Fig.\ 4.
A diagram contributing to $\bra\Om T J J \ket\Om$ at order $1$ involving a
quark loop.
The cut of the diagram shown contains two color-singlet states, and so can
contribute to 1-loop meson processes such as the one shown on the right.

\smallskip\noindent
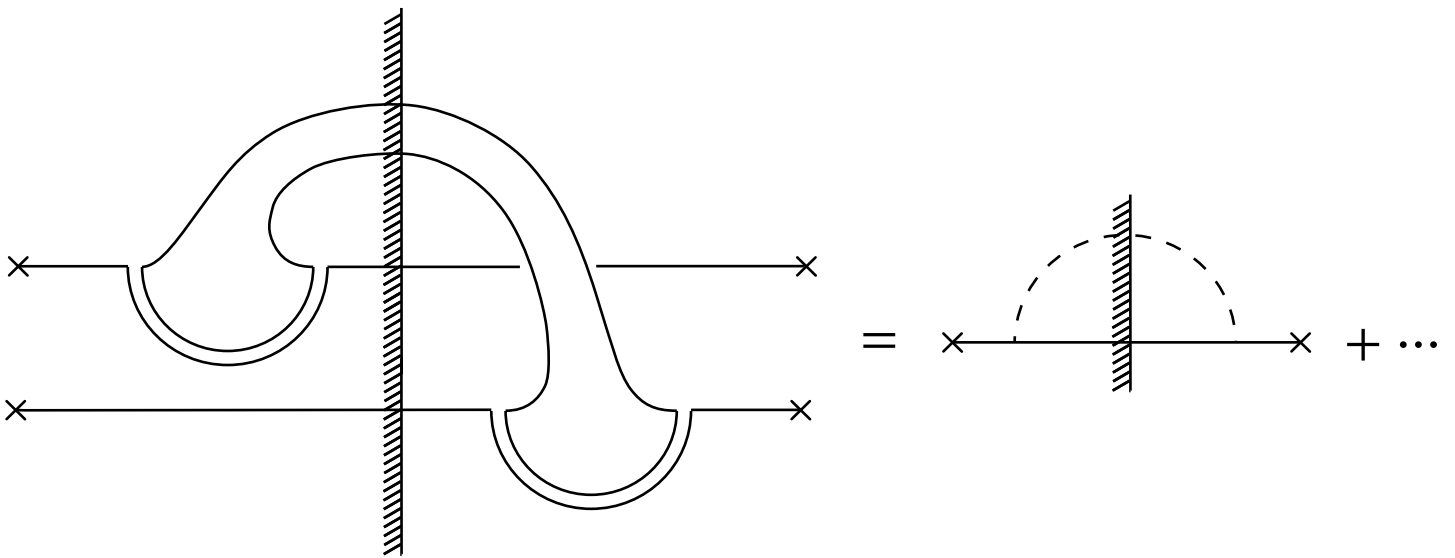
Fig. 5.
A diagram contributing to the baryon mass.
Counting factors of $N$ from the gluon vertices and color loops, it is easy
to see that this graph is $O(1/N)$ times a 2-body operator, and can
therefore be $O(N)$.
The cut of the diagram shown can contains 2 color-singlet intermediate
states:
the ``valence'' color lines which attach to the baryon state form a color
singlet with the remaining quarks in the baryon, and the quark-antiquark pair
at the top of the cut can be a color singlet.
This graph can therefore contribute to 1-loop meson-baryon processes such as
the one shown on the right.

\smallskip\noindent
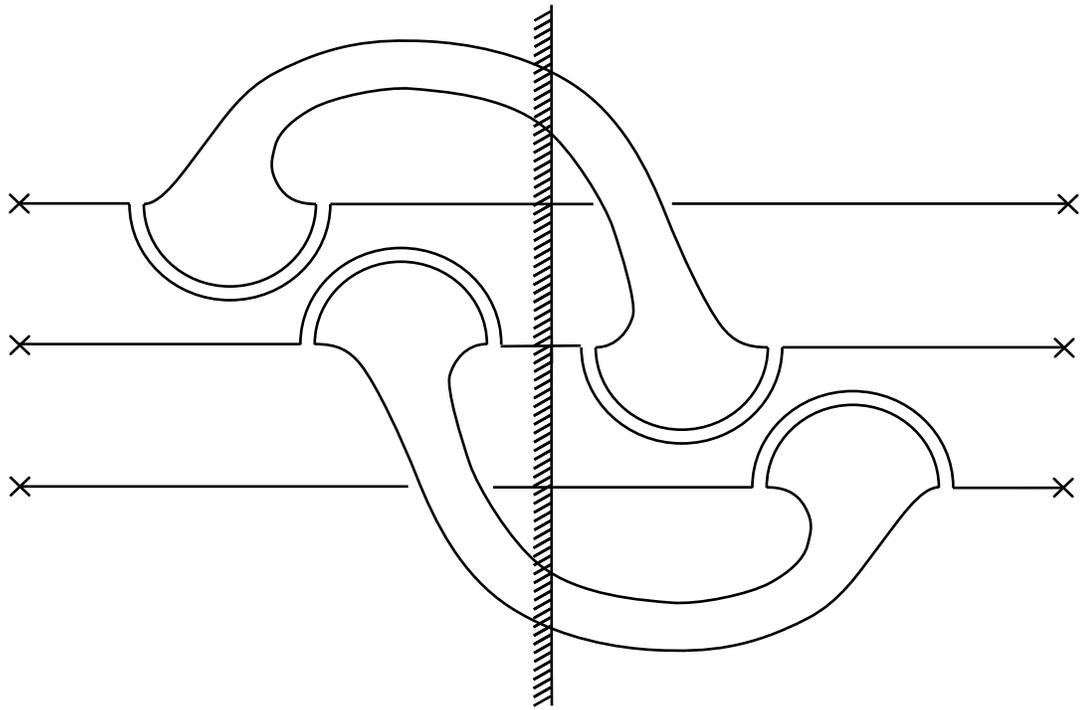
Fig. 6.
A diagram contributing to the baryon mass obtained by ``iterating'' the
diagram of fig.\ 5.
The cut shown contains 3 color-singlet intermediate states.

\smallskip\noindent
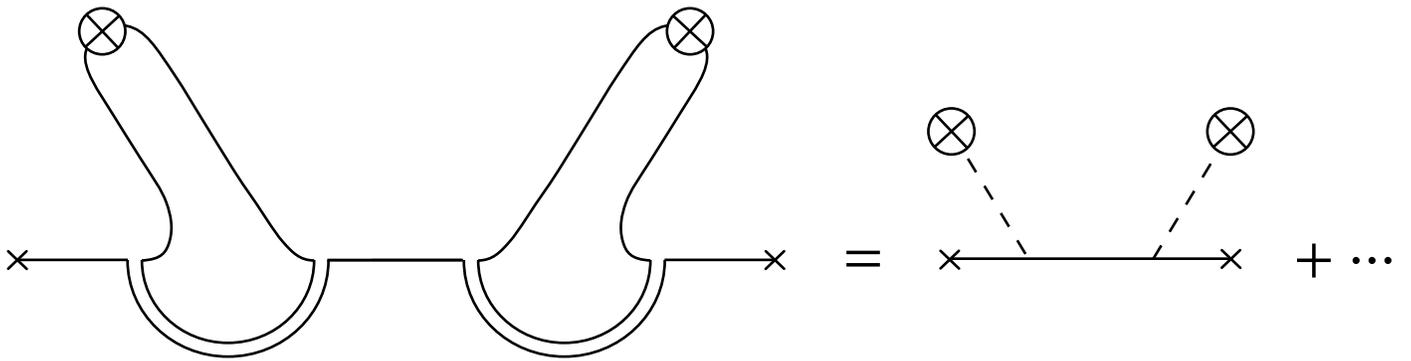
Fig. 7.
A diagram contributing to the matrix element
$\bra{\scr B} T J J \ket{\scr B}$.
This graph is $O(1)$ times a 1-body operator, and can therefore be
$O(N)$.
It clearly has the right intermediate-state structure to contribute to the
baryon-meson scattering process shown on the right.

\smallskip\noindent
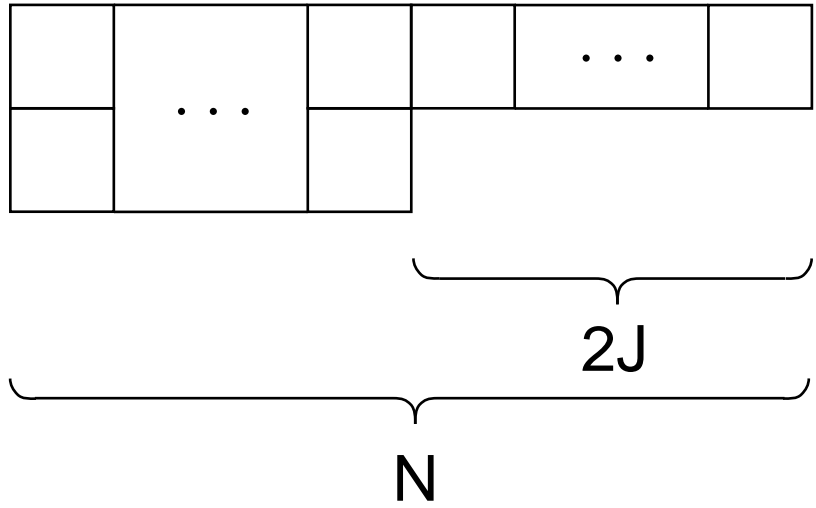
Fig. 8.
The Young tableaux corresponding to the $SU(N_F)$ representation for
baryons of spin $J$.

\smallskip\noindent
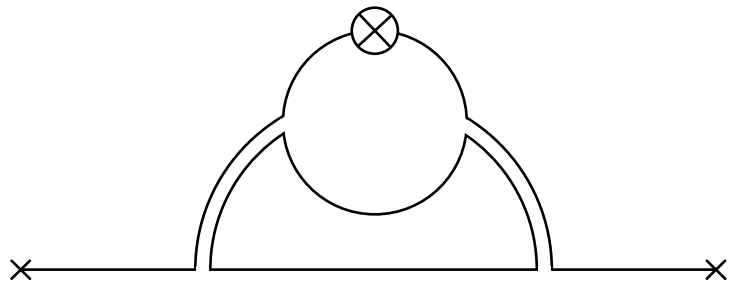
Fig. 9.
A leading diagram contributing to the matrix elements
$\bra p \mybar s \gam_\mu \gam_5 s \ket p$ or $\bra p \mybar ss \ket p$.

\bye